\documentclass[%
 reprint,
 amsmath,amssymb,
 aps,
]{revtex4-2}

\usepackage{graphicx}
\usepackage{dcolumn}
\usepackage{bm}

\begin{document}


\title{Ising-machine-like Driving Condition Restrictions in SSBM and Additional Pseudo-annealing\\- Driving Method for SSBM that Achieves Best Cut Value Search -} 



\author{Toshiya Sato}
\affiliation{NTT Device Technology Laboratories, NTT, Inc., Atsugi, Kanagawa 243-0198, Japan}


\date{\today}

\begin{abstract}
In our previous work, we proposed and experimentally demonstrated a physics-based simulator for combinatorial optimization problems (SSBM), which exploits a distinctive spontaneous symmetry breaking (SSB) phenomenon identified in a dissipative system exhibiting robust causality described by an iterative equation. We further conducted numerical simulations in which initial fluctuations of distinct patterns were applied to large-scale benchmark problems, and reported that the SSBM exhibits a unique behavior absent in conventional Ising machines: under specific conditions, pseudo-spin patterns progressively converge during the solution-search process and ultimately coalesce into a single pattern, with the cut value of the converged solution reaching 99.7\% of the best-known cut value. In the present paper, we elucidate the conditions under which the SSBM can be configured to emulate the behavior of a continuous-variable Ising machine. We further demonstrate that by additionally incorporating an effect analogous to physical annealing into these conditions, the SSBM becomes capable of identifying the best-known cut value for the K2000 benchmark instance.
\end{abstract}

\pacs{}

\maketitle 

\section{Introduction}

The realization of high-performance solvers for combinatorial optimization problems is a critically important challenge driven by real-world demands—such as drug discovery, logistics, and portfolio optimization—that require the selection of the most suitable option from an enormous number of candidates. Following the proposal and physical implementation of a quantum annealing-based Ising machine known as D-Wave \cite{Nishimori,Johnson}, an extended Ising machine capable of handling fully connected problems was subsequently realized through a physically implemented Ising machine based on a ring-cavity pulsed laser oscillator system \cite{Yamamoto}. Moreover, propelled by the remarkable advances in digital processing technologies that leverage GPUs and FPGAs—platforms that offer substantial advantages in parallel computation—noteworthy achievements have also begun to emerge in the category of algorithm-based solvers, for which parallelization-induced acceleration plays a pivotal role. Representative examples include the physically inspired, ordinary differential equation-based SBM \cite{Goto}, simulated CIM, and various annealers designed to simultaneously achieve full connectivity and high-speed parallel processing, such as the Digital Annealer \cite{Aramon}, Momentum Annealing \cite{Okuyama}, and STATICA \cite{KYamamoto}.

Among these approaches, CIM, SBM and simulated CIM — which do not belong to the annealer category — share with the Hopfield–Tank model \cite{HopfieldT} the common feature of employing continuous variables (i.e., spin-corresponding variables that take analog values) within a continuous-time framework. At the same time, these methods possess the following two characteristics that are absent from the Hopfield–Tank model. First, (1) they map spins onto continuous variables and employ a process in which a pitchfork bifurcation drives a single stable equilibrium point toward one of two ultimate stable equilibria. Second, (2) to suppress unintended effects of the Ising interaction terms arising from the introduction of continuous variables, a steady-state approximation or the adiabatic theorem is assumed, and the bifurcation process is carried out sufficiently slowly \cite{Yamamoto, Goto}. Consequently, except during the initial phase, state transitions in the solution-search process occur while most of the continuous variables corresponding to spins remain in the vicinity of gradually growing stable equilibrium points; only a subset of continuous variables undergoes rapid spin-state flipping.

In contrast, we have independently proposed a novel type of physics-based simulator for combinatorial optimization problems, designated SSBM, grounded in the physical implementation of spontaneous symmetry breaking (SSB) \cite{Anderson, Onsager, Higgs}, and have successfully demonstrated its operation experimentally \cite{Toshiy01} (see also Section S1 and Fig. S1 in the Supplemental Material). Furthermore, through numerical characterization of the SSBM on the large-scale benchmark problem $K_{2000}$\cite{Yamamoto-2}, we reported the discovery of operating conditions under which the system exhibits a unique behavior absent in conventional Ising machines: pseudo-spin patterns progressively converge during the solution-search process and ultimately coalesce into a single pattern whose cut value reaches 99.7\% of that of the best-known solution \cite{GOTO-02} (hereinafter referred to as the converged solution). Nevertheless, the SSBM had not yet succeeded in identifying the best-known solution for $K_{2000}$. In the present paper, we report the discovery of conditions under which the best-known solution can be identified, achieved through a combined approach in which the solution-search conditions of the SSBM are tuned to replicate effects analogous to the steady-state approximation and adiabatic theorem employed in CIM and SBM, with the additional incorporation of an effect analogous to physical annealing.

\section{SSBM as a Fully Dissipative Connection System}

\begin{figure}[htbp]
  \centering
  \includegraphics[width=0.48\textwidth]{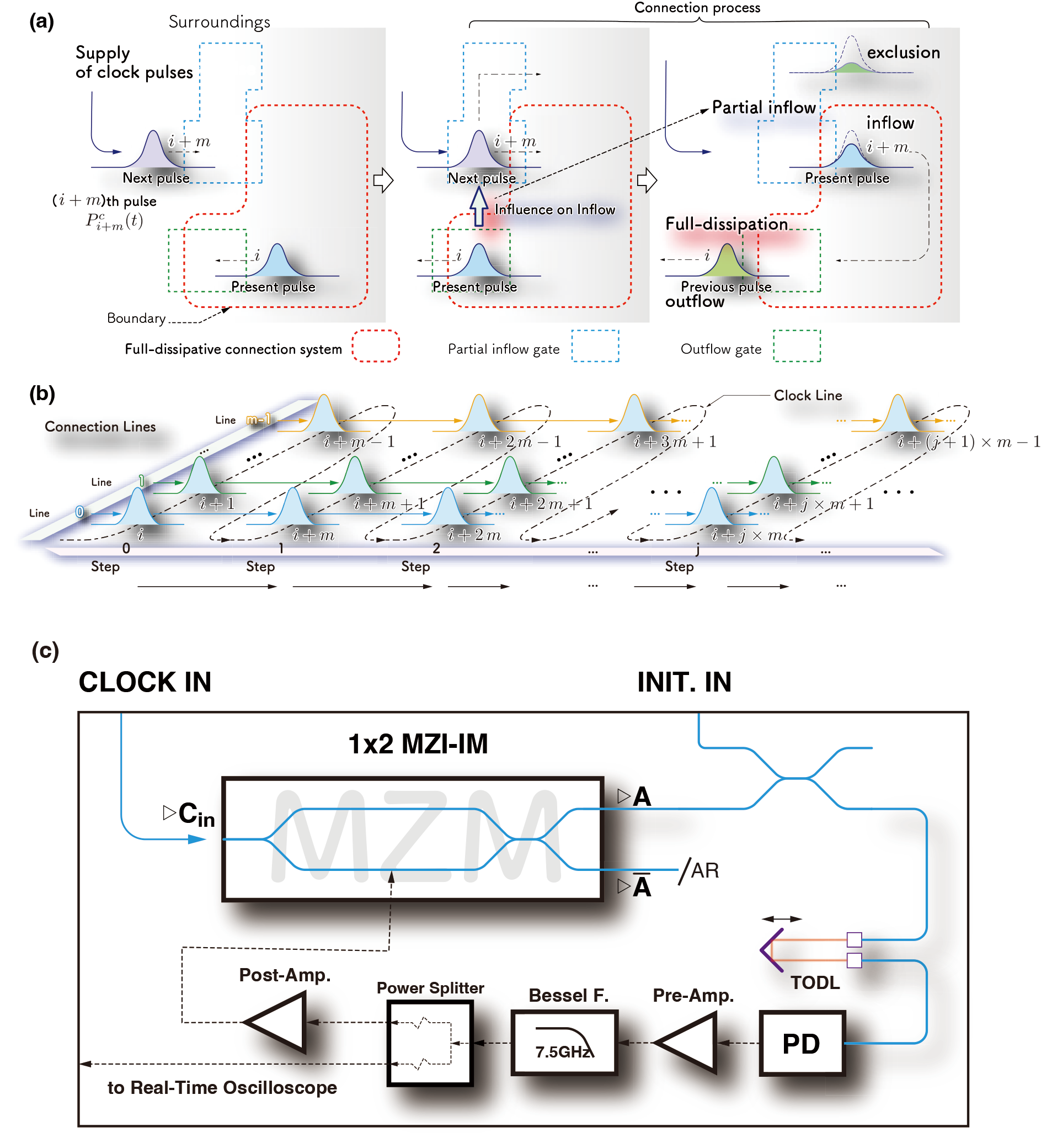} 
  \caption{Conceptual diagrams of the fully dissipative connection system: (a) the fundamental element system, (b) inter-pulse connections in time-division multiplexing, and (c) the photonic circuit implementing this system. The boundary of the fundamental element system (a) is not a simple spatial boundary but a spatiotemporal domain encompassing only the influencing pulse and the pulse that receives its influence. Sufficient isolation between individual clock pulses clearly defines this boundary, which is successively concatenated and parallelized through time-division multiplexing, collectively constituting $m$ independent fully dissipative connection systems. In this system, each incoming optical clock pulse simultaneously controls the transmittance of the subsequent pulse and dissipates completely out of the system. Within the photonic circuit, a $1\times2$ MZM functions as a partial-transmittance input gate, controlled via optoelectronically converted input pulses with optimized timing, pulse width, and output intensity.}
  \label{fig:01}
\end{figure}

The SSBM is a physics-based simulator for combinatorial optimization problems that exploits the SSB phenomenon, and is composed of $N$ fundamental element systems — equal in number to the problem size $N$ — each of which constitutes a dissipative rather than a Hamiltonian system (Fig. 1). In this system, an electrical pulse derived from a preceding incoming optical clock pulse drives the input gate of the system, thereby controlling the transmittance of the subsequent optical clock pulse while simultaneously dissipating completely out of the system. Furthermore, by satisfying the condition $\sigma_{\mbox{\rm \scriptsize pw}} \ll \tau_{\mbox{\rm \scriptsize pw}} \ll \ Delta t$ denotes the pulse width of the optical clock pulse, the modulation pulse width driving the input gate, and the repetition period of the optical clock pulse, respectively — a Mach–Zehnder interferometer-type intensity modulator (MZM) with a sufficiently wide response bandwidth, employed as the input gate, causes each fundamental element system to exhibit robust causality, referred to here as dissipative causality, described by the following iterative equation:

\begin{eqnarray}
\phi_{i+m} 
& = &
\sin^{2}(\frac{\gamma}{2} \, \phi_{i}  +  \theta_{B} ) 
\end{eqnarray}

where $\gamma$, $\phi_{i+m}$ and $\theta_{B}$ denote the modulation efficiency, the transmittance to port ${\bf A}$ and the static phase condition of the MZM, respectively. The multiple dissipative systems constituting the fundamental element systems of the SSBM can be realized not only through spatial parallelization but also through time-division multiplexing. In the latter case, boundaries conventionally defined in the spatial domain are instead defined in the spatiotemporal domain \cite{Toshiy01} (see Figs. 1(a) and 1(b)).

In what follows, we further restrict our attention to the specific conditions $\gamma \!=\! \pi$ and $\theta_{B} \!=\! 0$. Under these conditions, the system is governed by a symmetric double-well pseudo-potential possessing one unstable fixed point ($\phi = 1/2$) with a symmetry axis passing through it, along with two attractors ($\phi = 0$ and $\phi = 1$). Upon initiation of the optical clock pulse input, the onset of dissipative causality inevitably gives rise to a symmetry-broken state in which the system is locked to the attractor $\phi = 0$. This state corresponds to a stable fixed point at which no clock pulse enters the system interior through the boundary under consideration, and conceals an important characteristic that is easily overlooked—hereafter referred to as the hidden state (HS). The symmetry already broken in the HS is restored by driving the system from the HS toward the unstable fixed point while preserving the aforementioned double-well pseudo-potential (see Section S1 of the Supplemental Material). In the time-division multiplexing implementation, this restoration is achieved by forcibly rewriting the system state via external input pulses: specifically, $N$ optical pulses—equal in number to the fundamental elements and with intensity adjusted to half that of the optical clock pulse — are injected through an external input port additionally provided between the input gate and the MZM, with the injection timing appropriately synchronized. The restored symmetry is then immediately and spontaneously broken by fluctuations, triggering a transition to one of the two stable fixed points via dynamics-derived attraction (DDA) — a process that constitutes the SSB phenomenon. A delayed optical interferometer is further employed to introduce inter-element coupling described by Eqs. (2)–(8), namely, pseudo-spin interaction (pSI).

\begin{widetext}
\begin{eqnarray}
\phi_{i+m}
& = &  {\cal C}( \,\frac{\pi}{2}, \,\phi_{i},  \,{\cal Q}_{i} \,)\nonumber\\
& = &
\left\{\sin^{2}(\frac{\pi}{2} \,|\sqrt{\phi_{i}} -  {\cal Q}_{i} |^{2} ) +  \overline{ \sin^{2}(\frac{\pi}{2} \,|\sqrt{\overline{\phi_{i}}} -  {\overline{\cal Q}_{i} } |^{2} )} \right\}/ \,2  \\
{\cal Q}_{i}
& = &  
{\cal Q}_{i}^{\rm FM} + {\cal Q}_{i}^{\rm AFM} \\
{\cal Q}_{i}^{\rm FM}
& = &  
- \sum_{i \ne k} {\cal J}_{i:k}^{\rm FM}( \sqrt{\phi_{i}}  -  \sqrt{\phi_{i+k}})  \\
{\cal Q}_{i}^{\rm AFM}
& = &  
- \sum_{i \ne k} {\cal J}_{i:k}^{\rm AFM}( \sqrt{\phi_{i}}  -  \sqrt{\overline{\phi_{i+k}}})
\end{eqnarray}
\begin{eqnarray}
 {\cal J}_{i:k}^{\rm FM} = \left\{ \begin{array}{ll}
  {\cal J}_{i:k} (>0) & \mbox{if} \mbox{\footnotesize \rm \,\,\,\,\,\,\,\,\,ferromagnetic\,}\\
                                 0 & \mbox{if}\,\,\, \mbox{\footnotesize \rm  anti-ferromagnetic\,} \end{array} \right. \\
 {\cal J}_{i:k}^{\rm AFM} = \left\{ \begin{array}{ll}  0 & \mbox{if}\,\,\, \mbox{\footnotesize \rm  \,\,\,\,\,\,\,\,\,ferromagnetic\,} \\
                                  {\cal J}_{i:k} (>0) & \mbox{if} \,\,\,   \mbox{\footnotesize \rm  anti-ferromagnetic\,}  \end{array} \right. 
\end{eqnarray}
\end{widetext}

Here, $\overline{\phi}$ denotes the complement of $\phi$ and can be implemented using the complementary output of the MZM. The rationale for averaging using conjugate terms is to counteract the tendency of pseudo-spins to accumulate in the zero state due to ferromagnetic-type pseudo-spin interactions, thus improving overall performance \cite{Toshiy02}. Although this increases the complexity of the system, the physical implementability of the architecture is preserved.

The pSI introduced here is fundamentally distinct from the Ising interaction in that it is capable of producing effects consistent with the interpretation of exchange interaction energy across the entire value range of the continuous variable $\phi$. As a consequence, cases arise in which SSB cannot be completed due to competition between the pSI effect (i.e., the aggregate pSI acting on individual elements) and DDA. This situation occurs specifically when the coupling is strong in problems involving a large number of elements under full connectivity, such as $K_{2000}$, where the competition between pSI and DDA causes the pseudo-spins to become detached from their respective attractors. This characteristic of pSI also gives rise to a solution-convergence effect that does not appear in continuous-variable Ising machines (IMs), and it has been confirmed that conditions exist under which a single high-quality solution can be obtained by appropriately narrowing the operating parameters \cite{Toshiy02}. To address this competition-induced failure of SSB completion, a physically implementable modified system \cite{Toshiy02} governed by the following equations is introduced to relatively suppress the pSI effect and thereby ensure the convergence of pseudo-spins to their respective attractors by enhancing the DDA effect:

\begin{widetext}
\begin{eqnarray}
\phi_{i+m}
& = &
{\cal N}(\,n, \,\pi, \, {\cal C}( \,\frac{\pi}{2}, \,\phi_{i},  \,{\cal Q}_{i} \,))\\
{\cal N}(n, \pi, P_{in} )
& := &
\underbrace{
\sin^{2}(\frac{\pi}{2}\,\sin^{2}(\cdots \frac{\pi}{2}\,\sin^{2}(}_{n}\frac{\pi}{2}\,P_{in} )\cdots))
\end{eqnarray}
\end{widetext}

Here, $n$ denotes the number of nested hierarchical levels (i.e., the number of times the function is composed), and a higher value of $n$ yields a stronger attraction toward the attractors.

\section{Restrictive Driving Conditions in SSBM and Additional Pseudo-Annealing Effects}

In our previous study, benchmark tests on the K2000 problem via numerical simulation confirmed that SSBM exhibits a unique behavior of converging to a single high-quality solution; however, the known best solution could not be obtained. To resolve this limitation, we focused on two key differences between SSBM and the improved SBM variants — dSBM, HbSBM and HdSBM — for which successful identification of the known best solution has been reported \cite{GOTO-02}.
The first distinction concerns the quasi-adiabatic operation common to continuous-variable Ising machines, including SBM variants, in which the bifurcation parameter (referred to as "gain" in the CIM context) is swept quasi-statically so that the computation proceeds while maintaining a near-quasi-stationary state \cite{Yamamoto,Goto}. This quasi-stationary maintenance ensures that the dynamics of the spin variables (continuous variables) faithfully follow the bifurcation of the attractors, thereby mitigating artifacts that arise when Ising interactions—originally defined for binary spin systems—are directly applied to systems with continuous-variable spin assignments (see Supplementary Materials: Section S-2). Accordingly, it is anticipated that operating SSBM under conditions that prevent the continuous variables from detaching from their attractors would bring its operational state closer to a quasi-stationary regime. As a trade-off, however, such conditions are also expected to forfeit the advantage of converging to a single high-quality solution, causing statistical variability across trials to become manifest.
The second point concerns the improvement method adopted in HbSBM to enable identification of the known best solution. This method was derived by starting from the application of the Nosé–Hoover method—originally formulated under the assumption of equilibrium conditions—to bSBM and subsequently extending it beyond the strict scope of that framework. It has been argued that the resulting "heating process" is the primary driver of the observed improvement. We propose revised driving conditions for SSBM developed through stepwise examination of these two points, and verify their effects via numerical simulation.

\begin{figure}[!t]
\centering
\includegraphics[width=0.48\textwidth]{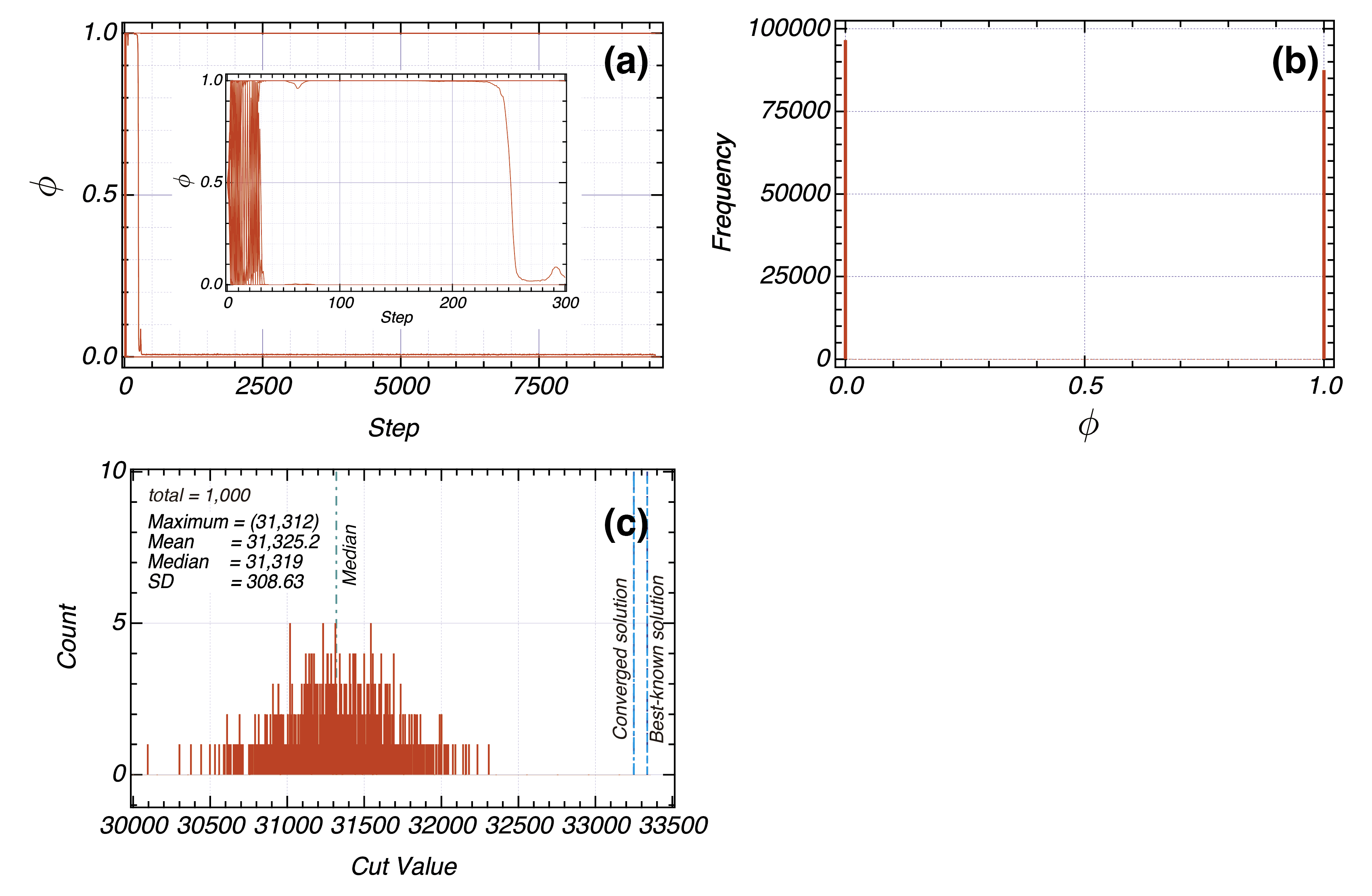}
 \caption{Behavior of SSBM under a restriction condition following the continuous-variable-type Ising machines (simulation results for the $K_{2000}$ problem, $n=12$, ${\cal J}_{i:k} = 4.985 \times 10^{-4}$). (a) Step evolutions of continuous-variables $\phi_{i}$ (pseudo-spins) (20 samples, 9699 steps), (b) Histogram of pseudo-spin values (20 samples, total=193,980 counts), (c) Histogram of cut values explored in 1,000 trials.}
 \end{figure}
\begin{figure}[!t]
\centering
\includegraphics[width=0.48\textwidth]{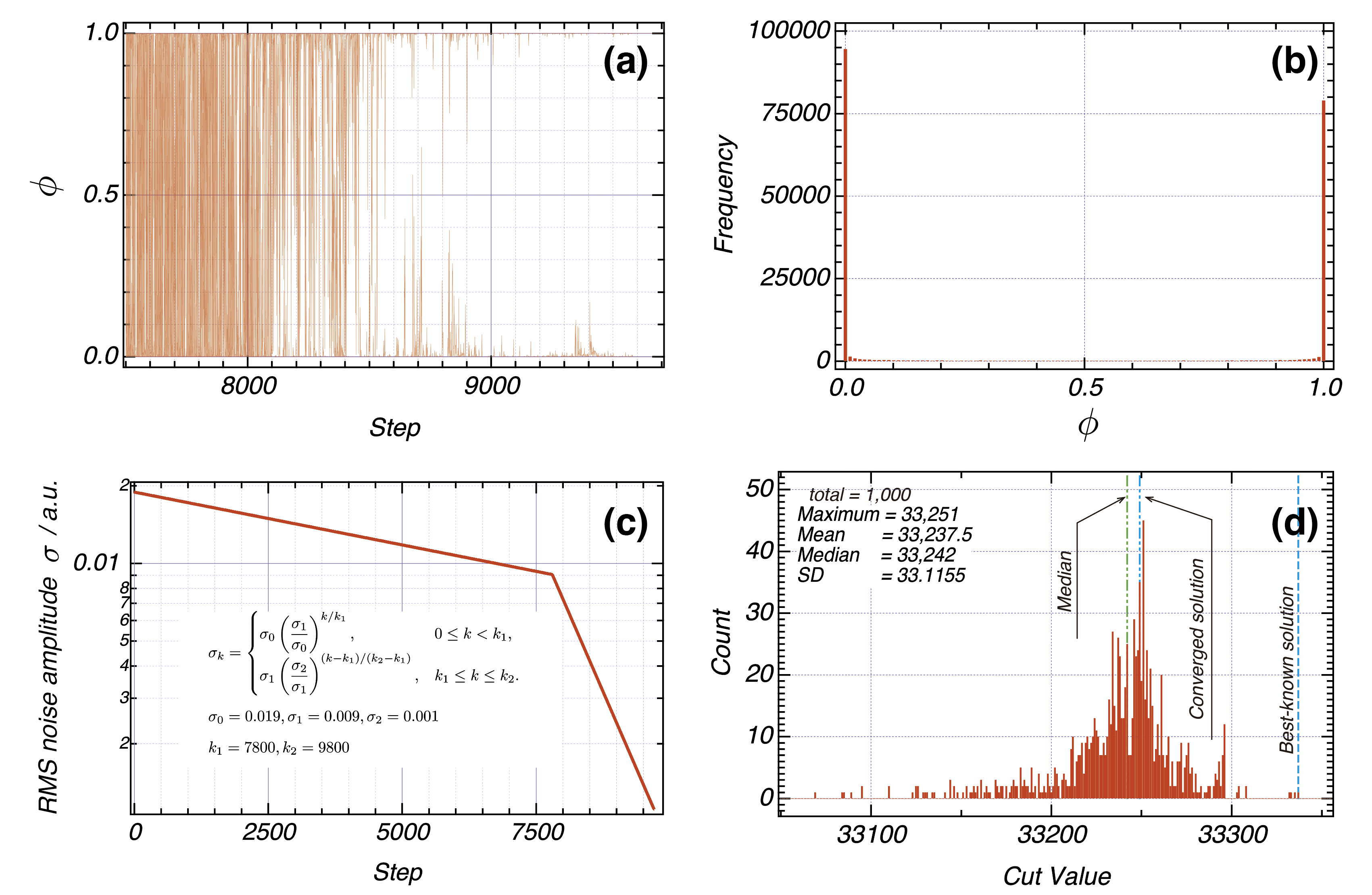}
 \caption{Behavior of SSBM in the case where a pseudo-annealing effect is added under the restricted conditions (simulation results for the $K_{2000}$ problem, $n=12$, ${\cal J}_{i:k} = 4.985 \times 10^{-4}$). (a) Step evolutions of pseudo-spins (20 samples, 9699 steps), (b) Histogram of pseudo-spin values (20 samples, total=193,980 counts), (c) Scheduling of additional noise levels for pseudo-annealing, (d) Histogram of cut values explored in 1,000 trials.}
 \end{figure}

Figure 2 presents plots of the pseudo-spin dynamics and the final cut values obtained from simulations performed to address the first point alone. The parameter $n$ was fixed at $n=12$ (see Supplementary Materials, Fig. S3), a value chosen to balance two competing requirements: maintaining pseudo-spins sufficiently attracted to their respective attractors while avoiding an extreme reduction in the flexibility of their continuous-variable dynamics. The simulations confirm two key findings. First, the pseudo-spins are sufficiently strongly confined near one of the two attractors, as intended (see Fig. 2(a),(b)). Second, the cut values obtained across trials initialized with different noise seeds exhibit a statistical distribution analogous to that observed in continuous-variable Ising machines (see Fig. 2(c)). This statistical spread in cut values—absent under the pseudo-spin pattern convergence conditions reported previously \cite{Toshiy02} — is attributed to the mutual isolation of individual solution-search trajectories, a consequence of the relatively large DDA that reduces the flexibility of pseudo-spin transitions.

Regarding the second point, two fundamental obstacles preclude the direct adoption of the HbSBM approach. First, the Nosé–Hoover method is a technique developed for continuous-time (differential) dynamical systems, whereas SSBM operates as a discrete dynamical system; therefore, the method cannot be applied as-is. Second, since HbSBM ultimately requires venturing beyond the foundational assumptions of the Nosé–Hoover method, clear and principled guidelines for its introduction remain ambiguous.
To circumvent these obstacles, we pursued an approach that differs from that employed in HbSBM and is directly implementable within SSBM. Specifically, by reference to the behavior of the instantaneous temperature reported in \cite{GOTO-02}, we investigated a pseudo-annealing scheme based on a piecewise exponential decay schedule \cite{Ingber1993, FraustoSolis2016}—in which the noise intensity gradually decreases through most of the computation before transitioning to a relatively rapid decay in the final stage to freeze the solution—thereby capturing the essence of the improvement effect achieved in HbSBM.

Figure 3 shows the results obtained with pseudo-annealing applied under the fixed condition $n=12$. The pseudo-spins are confirmed to remain strongly confined near their respective up and down attractor values while undergoing transitions between these states at a frequency governed by the instantaneous noise intensity (Fig. 3(a),(b)). In this simulation, Gaussian-distributed random noise was generated using the Mersenne Twister pseudorandom number generator \cite{Matsumoto1998}. The noise was injected according to the update rule $\sqrt{\phi_{i:k}} \mapsto \sqrt{\phi_{i:k}} + \sigma_k\,\xi_k$, where $\xi_k$ denotes a unit-variance random variable and $\sigma_k$ is the root-mean-square amplitude (standard deviation) scheduled as shown in Fig. 3(d). The range of $\sigma_k$ was adjusted to span from a value sufficiently large to render the effect of pSI negligible, down to a value sufficiently small to be itself negligible relative to the pSI effect. The distribution of cut values obtained from 1,000 trials (Fig. 3(d)) is confirmed to be shifted toward higher cut values compared with that obtained without pseudo-annealing (Fig. 2(c)).
This improvement can be understood as follows. During pseudo-annealing, noise-driven random transitions introduce entropic effects; simultaneously, the pSI stabilization effect—which had been suppressed by the confinement imposed under Point 1—is partially released. These two effects compete throughout the annealing process, and as the noise intensity decreases sufficiently according to the prescribed schedule, the pSI stabilization effect progressively becomes the dominant contribution.
Furthermore, in addition to the qualitative similarity between this cut-value distribution and those reported for dSBM, HdSBM and HbSBM (see Supplementary Materials, Fig. S4) — all of which have been shown to yield the known best solution \cite{GOTO-02} — the best cut value $(33,337)$ was also confirmed to be attained in the present trials. This similarity suggests that the two points highlighted in this study constitute conditions that exert a substantial influence on the solution-search behavior. Moreover, two additional observations warrant note. First, the median of the cut value distribution (33,242) was found to be close to the cut value of the previously reported converged solution (33,249)—defined here as the solution associated with a unique pseudo-spin pattern. Second, among the trials in this study that yielded a cut value of 33,249—identical to that of the previously reported converged solution—(35 out of 1,000 trials), 25 distinct pseudo-spin patterns were observed, in contrast to the single pattern found in the previous report. This discrepancy is attributed to the difference in the nature of the underlying transitions—noise-driven random transitions in the present case versus smooth transitions that exploit the intrinsic characteristics of pseudo-spin interactions in the previous case—which gives rise to distinct manifestations of the entropic effect on the stabilization process.

\section{Conclusion}

In this paper, we discuss driving conditions for the SSBM that incorporate a pseudo-annealing scheme inspired by the behavior of the instantaneous temperature in SBMs that have successfully identified the best-known solution \cite{GOTO-02}. By emulating the spin dynamics of continuous-variable Ising machines and supplementing this with the pseudo-annealing scheme, it was confirmed that the SSBM becomes capable of identifying the best-known solution, albeit at the cost of losing the ability to identify the converged solution. It should be noted that while the driving conditions of the SSBM can be configured to emulate those of continuous-variable Ising machines, the converse does not hold: the driving conditions of continuous-variable Ising machines cannot be adapted to satisfy the conditions required for the SSBM to obtain the converged solution. Furthermore, it was confirmed that the median of the cut value distribution obtained under the driving conditions under which the best-known solution was identified in the SSBM is in close proximity to the cut value of the converged solution. This observation suggests that the phenomenon by which the SSBM searches for the converged solution involves, in certain cases, an effect that draws the system toward the converged solution in a manner that counteracts the energy stabilization effect. A detailed investigation of this effect remains a subject for future work.

\section*{Acknowledgments}
The authors thank T. Hashimoto and H. Takenouchi for their support.



%
%

%


\bibliography{your-bib-file}

\end{document}